%% file: main.tex
\documentclass[preprint,12pt]{elsarticle}

\usepackage{amsfonts}
\usepackage{amsmath}
\usepackage[toc,page]{appendix}
\usepackage{color}
\usepackage{epsfig}
\usepackage{graphics}
\usepackage{subfig}
\usepackage{amssymb}

\usepackage{lineno,hyperref}
\modulolinenumbers[5]

\usepackage{float}

\usepackage{graphicx}
\usepackage{amssymb}

\usepackage{ulem}

\usepackage{lineno}

\journal{Elsevier - Expert System with Applications}

\begin{document}

\begin{frontmatter}

\title{Complex Correntropy Function: properties, and application to a channel equalization problem}








\author{Jo\~ao P. F. Guimar\~aes, Aluisio I. R. Fontes, Joilson B. A. Rego, Allan de M. Martins, J.C. Principe}

\address{Federal Institute of Rio Grande do Norte, 59015-300, Natal, Brazil\\
e-mails: \{joao.guimaraes,aluisio.rego\}@ifrn.edu.br}

\address{Department of Computer Engineering and Automation\\
Federal University of Rio Grande do Norte, 59078-900, Natal, Brazil\\
e-mails: \{allan,\,joilson\}@dca.ufrn.br}

\address{Department of Electrical and Computer Engineering\\
University of Florida, Gainesville, FL 32611 USA \\
e-mails: \{principe\}@cnel.ufl.edu}

\begin{abstract}

The use of correntropy as a similarity measure has been increasing in different scenarios due to the well-known ability to extract high-order statistic information from data. Recently, a new similarity measure between complex random variables was defined and called complex correntropy. Based on a Gaussian kernel, it extends the benefits of correntropy to complex-valued data. However, its properties have not yet been formalized. This paper studies the properties of this new similarity measure and extends this definition to positive-definite kernels. Complex correntropy is applied to a channel equalization problem as good results are achieved when compared with other algorithms such as the complex least mean square (CLMS), complex recursive least squares (CRLS), and least absolute deviation (LAD).

\end{abstract}


\end{frontmatter}


\section{Introduction}

Recent studies have demonstrated that correntropy is an efficient tool for analyzing higher-order statistical moments in machine learning and signal processing applications \citep{liu2007correntropy,fontes2014classification, Fontes2015, fontes2017cyclostationary}. In particular, the maximum correntropy criterion (MCC) has been successfully used in filtering, robust regression, and signal processing applications. Correntropy extends the autocorrelation function to the nonlinear systems analysis and nonGaussian sources from real-valued data \citep{peng2017constrained,he2011maximum}.

On the other hand, many applications associated to signal processing involve signal sources that employ complex-valued data. However, correntropy has only been defined for real numbers \citep{Santamaria2006a}. So, due to the limitation to deal with real-valued data, it is difficult to use correntropy in a straightforward way as applied to problems involving complex-valued data. Inspired by the probabilistic interpretation demonstrated in \citep{integralxy}, our research group has extended the concept of classic correntropy to include processing from complex-valued data \citep{7763864}, as the method has been defined as complex correntropy. This similarity measure is based on the probability density function (PDF) applied to multidimensional spaces using the Parzen estimator with a Gaussian Kernel. Additionally, a new cost function called Maximum Complex Correntropy Criterion (MCCC) is defined, whose performance has proven to be superior to that regarding classical algorithms for processing complex-valued data \citep{7763864}. However, the properties of complex correntropy have yet not been properly formalized.

This paper extends the complex correntropy definition to positive-definite kernels and gives a probabilistic interpretation for the case of positive valued kernels. Besides, some important properties are presented and studied, which make complex correntropy structurally similar to the correntropy function. Complex correntropy is also used as a cost function in a channel equalization problem, while the results demonstrate the advantages of this new similarity function in nonGaussian environments when compared to LAD, CLMS, and CRLS algorithms.

The organization of the paper is as follows. Section \ref{PI} defines the complex correntropy, as some important properties are presented. In Section \ref{MCCC}, the maximum complex correntropy criterion (MCCC) is presented, where Wirtinger derivatives are briefly discussed and used to obtain a fixed-point recursive algorithm. Section \ref{results} describes a channel equalization problem used to illustrate and verify the theoretical assumptions associated to the properties of MCCC, which provide overall improved performance if compared with the classical solutions. Finally, Section \ref{conc} summarizes the main conclusions and potential future work.

\section{Definition and Properties of Complex Correntropy }
\label{PI}

This section presents the complex correntropy definition and properties. Firstly, let us assume that $C_1$ and $C_2$ are always complex random variables, where $C_{1}=X+j\,Z$ and  $C_{2} = Y+j\,S$ with $X;Y;Z;S$  being real-valued random variables; and $j$ refers to the imaginary unit.

\subsection{Definition}

Complex Correntropy is a generalized similarity measure between two arbitrary scalar complex random variables $C_{1}$ and $C_{2}$ defined as

\begin{equation}\label{correntropia_complexa}
V^{C}_{\sigma}(C_{1},C_{2}) =  E_{C_{1}C_{2}}[ K(C_{1},C_{2}) ]
\end{equation}

where $K$ is a positive-definite kernel.

\subsection{Properties:}

Important properties of the complex correntropy are listed and proved in this section. Since complex correntropy was developed to keep important characteristics from the classical correntropy, some properties such as symmetry, high-order statistical measure, probabilistic interpretation,  consistent and asymptotically unbiased estimator, are the same for both cases (i.e. properties 1, 3, 4 and 5). Other properties must be properly adapted and modified e.g. boundedness (property 2). Besides, two new properties exist in the case of complex correntropy: one of them connects two similarity measures numerically (property 6), while the other one is adequate for the polar representation of complex numbers (property 7).

\bigskip
\textit{Property 1:} For symmetric kernels, complex correntropy is symmetric.

\begin{equation}
V^{C}_{\sigma}(C_{1},C_{2}) = E_{C_{1}C_{2}}[ K (C_{1}, C_{2}) ]  = E_{C_{1}C_{2}}[ K(C_{2},C_{1}) ] = V^{C}_{\sigma}(C_{2},C_{1}) 
\end{equation}

\textit{Proof.} This property follows from the concepts of positive definiteness and symmetry. 
\begin{flushright}
$\blacksquare$
\end{flushright}

\bigskip
\textit{Property 2:} Complex correntropy is positive and bounded. For the Gaussian kernel, its estimated value $\hat{V}^C_{\sigma}$ is always real and between zero, the minimum value, and  $1 / 2\pi\sigma^2 $, which is achieved when $C_1 = C_2$.

\begin{equation}
0 \leq \hat{V}^{C}_{\sigma}(C_{1},C_{2}) \leq \frac{1}{2\pi\sigma^2}
\end{equation}

\textit{Proof.} Applying the Gaussian kernel $G^{C}_{\sigma}$, defined in (\ref{kernelcomplexogaussiano}), to the definition established by (\ref{correntropia_complexa}) gives (\ref{ccaberta}).

\begin{equation}\label{kernelcomplexogaussiano}
G^{C}_{\sigma} (C_1 - C_2 )= \frac{1}{2\pi\sigma^2}exp \left ( -\frac{(C_{1} - C_{2}) (C_{1} - C_{2})^{*}}{2\sigma^2} \right )
\end{equation}

\begin{equation}\label{ccaberta}
\hat{V}^{C}_{\sigma}(C_{1},C_{2}) = \frac{1}{2\pi\sigma^{2}} \frac{1}{N} \sum\limits_{n=1}^N exp \left ( -\frac{(x_{n} - y_{n})^{2} + (z_{n} - s_{n})^{2} }{2\sigma^2} \right ) 
 \end{equation}

If $C_1 = C_2$, $x = y$ and $z=s$, $\forall n$, the, $ \hat{V}^{C}_{\sigma}(C_{1},C_{2}) = 1 / 2\pi\sigma^2 $. Otherwise, if $C_1$ is much different from $C_2$, the negative exponential term of equation (\ref{ccaberta}) increases and $\hat{V}^C_{\sigma}(C_1,C_2)$ tends to zero, thus validating the property. $\blacksquare$

\bigskip
\textit{Property 3:} For the Gaussian kernel, the complex correntropy is a weighted sum of all the even moments of the random variable $C_1 - C_2$. Furthermore, increasing the kernel size makes correntropy tends to the correlation of $C_1$ and $C_2$.

\textit{Proof.} Recalling that $C_1 = X + jZ$ and $C_2 = Y + jS$, let us analyze equation (\ref{ccaberta}) according to its respective Taylor series expansion, which gives

\begin{equation}\label{tayloraberta}
\begin{split}
V_{\sigma}^C(C_1,C_2) = \frac{1}{2\pi\sigma}  E[ 1 - \frac{(X-Y)^2}{2\sigma^2}  + \frac{(X-Y)^4}{8\sigma^4} - \frac{(X-Y)^6}{48\sigma^6} + \frac{(X-Y)^8}{384\sigma^8} +\\
-  \frac{(Z-S)^2}{2\sigma^2} + \frac{(Z-S)^4}{8\sigma^4} - \frac{(Z-S)^6}{48\sigma^6} + \frac{(Z-S)^8}{384\sigma^8} + \frac{(X-Y)^2 (Z-S)^2}{4\sigma^4} + \\
- \frac{(X-Y)^2 (Z-S)^4}{16\sigma^6} - \frac{(X-Y)^4 (Z-S)^2}{16\sigma^6} + \frac{(X-Y)^2 (Z-S)^6 }{96\sigma^8}\\
+ \frac{(X-Y)^4 (Z-S)^4}{64\sigma^8} + \frac{(X-Y)^6 (Z-S)^2 }{ 96\sigma^8 } + ... ]
\end{split}
\end{equation}

One could group the terms with $\sigma^2$ in the denominator and define $h_{\sigma^4}$ as a variable containing the high-order terms of the summation. Then, it is possible to write

\begin{equation}\label{taylor}
V_{\sigma}^C(C_1,C_2) = \frac{1}{2\pi\sigma}   - \frac{1}{2\pi\sigma} E[ \frac{(X - Y)^2 + (Z-S)^2}{2\sigma^2}]  + h_{\sigma^4}
\end{equation}

\begin{equation}
V_{\sigma}^C(C_1,C_2) = \frac{1}{2\pi\sigma}   - \frac{1}{2\pi\sigma} E[(C_1 - C_2)(C_1-C_2)^* ]  + h_{\sigma^4}
\end{equation}

\begin{equation}
V_{\sigma}^C(C_1,C_2) = \frac{1}{2\pi\sigma}   - \frac{1}{2\pi\sigma} R[C_1,C_2]  + h_{\sigma^4}
\end{equation}
where $R[C_1,C_2] = E[C_1\,C_2^*]$ is the correlation between $C_1$ and $C_2$. 
One can notice in equation (\ref{taylor}) that the higher-order terms represented by $h_{\sigma^4}$ tend to zero faster than the second one as $\sigma$ increases, what corresponds exactly to the correlation involving two complex variables $C_1$ and $C_2$, which completes the proof.

\begin{flushright}
$\blacksquare$
\end{flushright}


\newcommand{\limit}[3]
{\ensuremath{\lim_{#1 \rightarrow #2} #3}}

\bigskip
\textit{Property 4:} When using the Gaussian Kernel with kernel size $\sigma$, decreasing $\sigma$ to zero causes complex correntropy to approach the value associated to the probability density of the event $C_1=C_2$ ($P(C_1=C_2)$). 

\begin{equation}
\limit{\sigma}{0} V^{C}_{\sigma}(C_{1}, C_{2}) = \int\limits_{-\infty}^{\infty} \int\limits_{-\infty}^{\infty} f_{XYZS}(u_1,u_1,u_2,u_2) \mathrm{d}u_1\mathrm{d}u_2 = P(C_1 = C_2)
\end{equation}
where $f_{XYZS}(u_1,u_1,u_2,u_2)$ is the joint PDF.

\textit{Proof:} 
Let us recall the complex correntropy definition presented in Equation (\ref{correntropia_complexa}) i.e. $V^{C}(C_{1},C_{2}) =  E_{C_{1}C_{2}}[ K(C_{1},C_{2}) ]$. Two complex numbers are equal when both their real and imaginary parts are also equal to each other. Then, using the Gaussian kernel and expanding the terms associated $C_1$ and $C_2$, one can obtain
\begin{equation}
V^{C}_{\sigma}(C_{1}, C_{2}) = \int\limits_{-\infty}^{\infty} \int\limits_{-\infty}^{\infty} \int\limits_{-\infty}^{\infty} \int\limits_{-\infty}^{\infty} f_{XYZS}(x,y,z,s) G_{\sigma}(x-y) G_{\sigma}(z-s) \mathrm{d}x\mathrm{d}y \mathrm{d}z\mathrm{d}s
\end{equation}

From the theory of distributions, the following statement can be demonstrated

\begin{equation}
\limit{\sigma}{0}{G_{\sigma}} (x) \equiv \delta(x)
\end{equation}

Then, it gives

\begin{equation}\label{integral4proba}
\limit{\sigma}{0} V^{C}_{\sigma}(C_{1}, C_{2}) = \int\limits_{-\infty}^{\infty} \int\limits_{-\infty}^{\infty} \int\limits_{-\infty}^{\infty} \int\limits_{-\infty}^{\infty} f_{XYZS}(x,y,z,s) \delta(x-y) \delta(z-s) \mathrm{d}x\mathrm{d}y \mathrm{d}z\mathrm{d}s
\end{equation}

Making $x=y=u_1$ and $z=s=u_2$ results in

\begin{equation}\label{integral4prob}
\limit{\sigma}{0} V^{C}_{\sigma}(C_{1}, C_{2}) = \int\limits_{-\infty}^{\infty} \int\limits_{-\infty}^{\infty} f_{XYZS}(u_1,u_1,u_2,u_2) \mathrm{d}u_1\mathrm{d}u_2 = P(C_1 = C_2)
\end{equation}

\begin{flushright}
$\blacksquare$
\end{flushright}

Complex correntropy was defined in order to maintain the probabilistic meaning of correntropy. Equation (\ref{integral4prob}) represents a variable transformation of the joint PDF, $f_{XYZS}$, which is evaluated with equal arguments $u_1$ and $u_2$. By using a Parzen method in \citep{parzen1962estimation} to estimate $f_{XYZS}$, when $\sigma$ goes to zero and the product $N \sigma$ tends to infinity, respectively, the estimative of the joint PDF, $\hat{f}_{XYZS}(x,y,z,s)$, approaches $f_{XYZS}(x,y,z,s)$ asymptotically in the mean square sense. Therefore, using the Gaussian Kernel with a small kernel size in the Parzen estimation of $f_{xyzs}$  causes Equation (\ref{integral4prob}) to provide a scalar value that approaches the probability density of the event $(C_1 = C_2)$. A thorough discussion on this probabilistic meaning is presented in \citep{7763864}.

\textit{Property 4.1:} Assuming i.i.d data $\{ ( x_i,y_i,z_i,s_i )_{i=1}^N \}$ draw from the joint PDF $f_{xyzs}$, while $\hat{f_{\sigma}}_{xyzs}$ is its respective Parzen estimate with kernel size $\sigma$, the complex correntropy estimated with kernel size $\sigma' = \sigma \sqrt{2}$ is the integral of $\hat{f_{\sigma}}_{xyzs}$ along the plane formed by $x=y$ and $z=s$.

\begin{equation}
\hat{V}^{C}_{\sigma'}(C_1,C_2) = \int\limits_{-\infty}^{\infty} \int\limits_{-\infty}^{\infty}  \! \hat{f_{\sigma}}_{XYZS}(x,y,z,s) \, \mathrm{d}u_{1}\mathrm{d}u_{2} \Big|_{x=y=u_{1}, z=s=u_{2}} 
\end{equation}

\textit{Proof.} For simplicity, the step-by-step proof is presented in detail in Appendix I.

\bigskip
\textit{Property 5:} Under the condition $N \rightarrow \infty$, $\hat{V}_N,\sigma(X,Y)$ is an estimator of $V_\sigma(C_1,C_2)$ consistent in mean square. Furthermore, under conditions $N\sigma \rightarrow \infty$ and $\sigma \rightarrow 0$, $\hat{V}_{N,\sigma}(C_1,C_2)$ is an asymptotically unbiased estimator for $p_E$ and consistent in mean square.

\textit{Proof.} Following the properties of the Parzen estimation \citep{parzen1962estimation}, one can say that, if the conditions

\begin{equation}
\lim_{n \rightarrow \infty} \sigma = 0;
\end{equation}
holds, them
\begin{equation}
\lim_{n \rightarrow \infty} \hat{f_{\sigma}}_{XYZS}(x,y,z,s) = f_{\sigma XYZS}(x,y,z,s)
\end{equation}
Thus, 
\begin{equation}
E[\hat{V}^C_{N,\sigma}(C_1,C_2) ] = V^C_\sigma (C_1,C_2)
\end{equation}
Recall the random variable $E = C_1 - C_2$ with PDF $p_E(e)$. By using the Gaussian kernel and holding the conditions $N \rightarrow \infty $ and $\sigma \rightarrow 0$, it is possible to write (see property 4)

\begin{equation}
\lim_{N \rightarrow \infty, \sigma \rightarrow 0} E[\hat{V}^C_{N,\sigma}(C_1,C_2) ] = P(C_1 = C_2) = p_E(0)
\end{equation}

Furthermore, to analyze the variance, it is also needed to consider the Parzen estimator properties shown on \citep{parzen1962estimation}, then

\begin{equation}
var[\hat{V}^C_{N,\sigma}(C_1,C_2) ] = N^{-1} var[G_{\sigma}(C_1 - C_2)]
\end{equation}

\begin{equation}
\lim_{N \rightarrow \infty, \sigma \rightarrow 0} N \sigma \, var[\hat{V}^C_{N,\sigma}(C_1,C_2) ] = p_{E}(0) \int G^2_{1}(u)du
\end{equation}

where $G_{1}(u)$ is the Gaussian Kernel with $\sigma = 1$.

\begin{flushright}
$\blacksquare$
\end{flushright}

\bigskip
\textit{Property 6:} For two real random variables, $R_1$ and $R_2 \in \mathbb{R}$, using the Gaussian Kernel, complex correntropy $V^{C}_{\sigma}(R_1,R_2)$ differs form conventional correntropy $V_{\sigma}(R_1,R_2)$ by a factor of $(\sqrt{2\pi}\sigma)$.


\begin{equation}\label{propriedade6}
\hat{V}^{C}_{\sigma}(R_1,R_2) \sqrt{2\pi}\sigma = \hat{V}_{\sigma}(R_1, R_2) 
\end{equation}

\textit{Proof.} Let us expand the left-hand side of equation \ref{propriedade6} as

\begin{equation}\label{igualdade2}
\hat{V}^{C}_{\sigma}(R_1,R_2) \sqrt{2\pi}\sigma = \frac{1}{2\pi\sigma^{2}} \frac{1}{N} \sum\limits_{n=1}^N exp \left ( -\frac{(x_{n} - y_{n})^{2} + (z_{n} - s_{n})^{2} }{2\sigma^2} \right ) \sqrt{2\pi}\sigma 
\end{equation}

If $R_1$ and $R_2$ do not have imaginary parts, one can write equation \ref{igualdade2} as

\begin{equation}
\hat{V}^{C}_{\sigma}(R_1,R_2) \sqrt{2\pi}\sigma = \frac{1}{\sqrt{2\pi}\sigma} \frac{1}{N} \sum\limits_{n=1}^N exp \left ( -\frac{(x_{n} - y_{n})^{2}  }{2\sigma^2} \right ) 
\end{equation}
which is exactly the expression to estimate $V_\sigma(R_{1},R_{2})$ using the Gaussian Kernel \citep{livroitl}. 
\begin{flushright}
$\blacksquare$
\end{flushright}

It is important to mention that, using the Gaussian kernel, one could expect complex correntropy to generalize the real correntropy also numerically, as to be equal to each other when the imaginary parts of the data be equal to zero. That is not the case because, as an estimate, the error in the estimation causes the Gaussians to "spreads" out differently in the 4 dimensional case (two complex variables) than it does in two.


\bigskip
\textit{Property 7:} Using the Gaussian kernel, the complex correntropy can be expressed in polar coordinates as

\begin{equation}
\hat{V}^{C}_{\sigma}(C_{1},C_{2}) = \frac{1}{2\pi\sigma^{2}} \frac{1}{N} \sum\limits_{n=1}^N exp \left ( - \frac{|C_1|^2 + |C_2|^2}{2\sigma^2}  - \frac{|C_1||C_2| cos(\theta-\phi)}{\sigma^2} \right)
\end{equation}

\textit{Proof.}
Since $C_1$ and $C_2$ can be written in the polar form as $C_1 = |C_1|cos(\phi) + j|C_1|sin(\phi)$, and $C_2 = |C_2|cos(\theta) + j|C_1|sin(\theta) $, equation (\ref{ccaberta}) can now be represented as

\begin{equation}\label{polar1}
\begin{split}
(x-y)^2 = (|C_1|cos(\theta) - |C_2|cos(\phi))^2 = \\
= |C_1|^2 cos(\theta)^2 -2|C_1||C_2|cos(\theta)cos(\phi) + |C_2|^2 cos(\phi)^2
\end{split}
\end{equation}

Then, it is possible to expand term $(z-s)^2$ as

\begin{equation}\label{polar2}
(z-s)^2 = (|C_1|sin(\theta) - |C_2|sin(\phi))^2 = |C_1|^2 sin(\theta)^2 -2|C_1||C_2|sin(\theta)sin(\phi) + |C_2|^2 sin(\phi)^2
\end{equation}

Now, let us make $(x-y)^2 + (z-s)^2 $ by adding equation \ref{polar1} and \ref{polar2} i.e.
\begin{equation}
|C_1|^2 + |C_2|^2 - 2|C_1||C_2|(cos(\theta)cos(\phi) + sin(\theta)sin(\phi)) = |C_1|^2 + |C_2|^2 - 2|C_1||C_2|cos(\theta-\phi)
\end{equation}

\begin{flushright}
$\blacksquare$
\end{flushright}

\section{Maximum Complex Correntropy Criterion}
\label{MCCC}

Complex correntropy has been defined as a robust similarity measure between two complex random variables \citep{7763864}. Let us consider a linear model and define the error $e = d - y$ as the difference between the desired signal $d$ and the estimated output $y = \textbf{w}^H \textbf{X}$, where $y$, $d$, $e$, $\textbf{X}$, $\textbf{w} \in \mathbb{C}$, and the superscript $H$ is the Hermitian (conjugate tranpose). Let the new criteria MCCC be defined as the maximum complex correntropy between two random complex variables $D$ and $Y$ \citep{7763864}.

\begin{equation}\label{CCJ3}
J_{MCCC} = V^{C}_{\sigma}(D, Y) = E_{DY}[G^{C}_{\sigma}(D-\textbf{w}^{H}\textbf{X})] =  E_{DY}[G^{C}_{\sigma}(e)]
\end{equation}
where $w$ is the parameter that controls the error between the estimated and the desired signal, as $G^{C}_{\sigma}$ is the Gaussian kernel.

One can obtain a fixed-point solution for the optimal weight by using equation (\ref{CCJ3}) as a cost function. Then, the natural way would be to set the cost function derivative to zero with respect to $\textbf{w}^*$. However, even depending on complex-valued parameters $(D,Y)$, as property 2 shows, using the Gaussian kernel makes complex correntropy a real-valued function. Then, equation (\ref{CCJ3}) is not analytical in the complex domain because the Cauchy-Riemann conditions are violated \citep{mandic2009complex}. Hence, standard differentiation can not be apply. One possible alternative to overcome this problem is to consider the cost function defined in an Euclidean domain with a double dimensionality $(\mathbb{R}^2)$, although this approach leads to onerous computations \citep{Bouboulis2011ExtensionLMS}. The Wirtinger Calculus, which will be briefly presented in this section later on, provide an elegant way to obtain a gradient of real-valued cost function that are defined in complex domains \citep{mandic2009complex,Bouboulis2011ExtensionLMS}.

\subsection{Wirtinger calculus}

The Wirtinger calculus was firstly introduced in \citep{Wirtinger1927}, based on the duality between spaces  $\mathbb{C}$ and $\mathbb{R}^{2}$. 

Let $f : \mathbb{C} \rightarrow \mathbb{C}$ be a complex function defined in C. Such function can also be defined in $\mathbb{R}^{2}$ ( i.e., $f(x + jy) = f(x,y)$).

The Wirtinger`s derivative of $f$ at a point $c$ is defined as follows \citep{Bouboulis2011ExtensionLMS}

\begin{equation}\label{eq1}
\frac{\partial f}{\partial z} (c) = \frac{1}{2} \left ( \frac{\partial f}{\partial x}(c) - j\frac{\partial f}{\partial y}(c) \right )
\end{equation}
while the Conjugate Wirtinger`s derivative of $f$ at $c$ is defined by
\begin{equation}\label{eq2}
\frac{\partial f}{\partial z^{*}} (c) = \frac{1}{2} \left ( \frac{\partial f}{\partial x}(c) + j\frac{\partial f}{\partial y}(c) \right )
\end{equation}

The work developed in \citep{Bouboulis2011ExtensionLMS} presents a complete overview on the properties associated to Wirtinger Calculus. In summary, in order to compute the Wirtinger derivative of a given function $f$, which is expressed in terms of $z$ and $z^{*}$, one should apply the usual differentiation rules after considering $z^{*}$ as a constant. Additionally, following the same concept to compute the Conjugate Wirtinger derivative of a function $f$, also expressed in terms of $z$ and $z^{*}$, one should apply the usual differentiation rules considering $z$ as a constant \citep{Bouboulis2011ExtensionLMS}. For example, considering $f$ as $f(z) = zz^{*}$ allows concluding that

\begin{equation}
\frac{\partial f}{\partial z} = z ^{*} \quad \text{and} \quad \frac{\partial f}{\partial z^{*}} = z
\end{equation}

\subsection{Fixed-point solution}

The MCCC fixed-point algorithm was firstly introduced and discussed in \citep{7763864}. By using Wirtinger calculus, it is possible to set the derivative of the cost function to zero i.e.

\begin{equation}\label{primeira}
\frac{\partial J_{MCCC} }{\partial w^{*}} = \frac{\partial E_{DY}[G^{C}_{\sigma\,\sqrt{2}}(e)] }{\partial w^{*}}  = E_{DY} \left [ G^{C}_{\sigma\,\sqrt{2}}(e)\frac{\partial (ee^*) }{\partial w^{*}} \right ]   =  \textbf{0}
\end{equation}

Which gives

\begin{equation}\label{derivada_w}
\frac{\partial (ee^*)}{\partial \textbf{w}^*} = \frac{\partial (d-\textbf{w}^{H}\textbf{X})(d^*-\textbf{w}^{T}\textbf{X}^*) }{\partial \textbf{w}^*} = (-d^*\textbf{X} + \textbf{XX}^{H}\textbf{w})
\end{equation}

Then, it is possible to substitute (\ref{derivada_w}) in (\ref{primeira}) as

\begin{equation}\nonumber
E_{DY}[G^{C}_{\sigma\,\sqrt{2}}(e) (-d^*\textbf{X} + \textbf{XX}^{H}\textbf{w}) ] = \textbf{0}
\end{equation}

\begin{equation}\label{deduzindo_passo3}
E_{DY}[G^{C}_{\sigma\,\sqrt{2}}(e)\,d^{*}\textbf{X}] = E_{DY}[G^{C}_{\sigma\,\sqrt{2}}(e)\textbf{XX}^{H}]\,\textbf{w}
\end{equation}

Finally, the final expression for $w$ is obtained as

\begin{equation}\label{finalpf}
\textbf{w} = \left [ \sum_{n=1}^{N} G^{C}_{\sigma\,\sqrt{2}}(e_{n})\textbf{X}_{n}\textbf{X}_{n}^{H} \right ]^{-1} \left [ \sum_{n=1}^{N} G^{C}_{\sigma\,\sqrt{2}}(e_{n})\textit{d}_{n}^{*}\,\textbf{X}_{n} \right ]
\end{equation}

A detailed step-by-step solution is presented in Appendix II.

\section{Simulations and results}
\label{results}

\begin{figure}[H] 
\centering
\input{teste.tex}
\caption{Block diagram representing a channel equalization problem.}
\label{figuraSistema}
\end{figure}
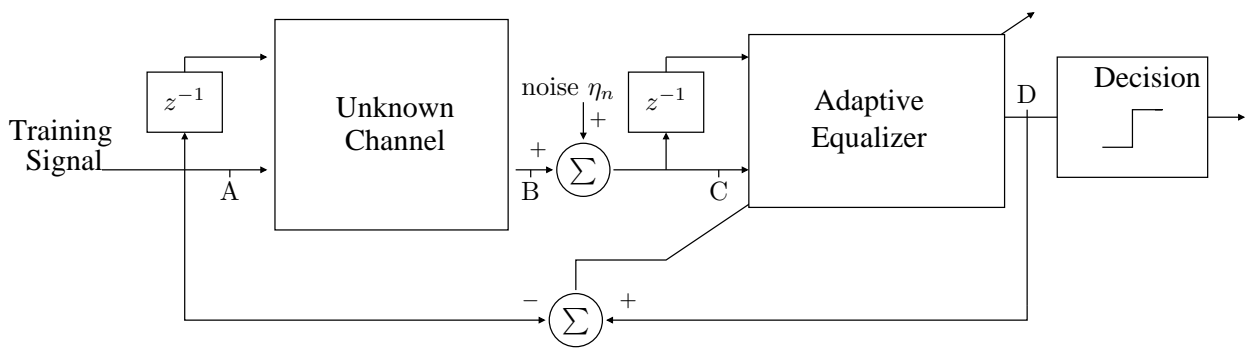

In order to evaluate the performance of the MCCC, a comparison has been established involving traditional algorithms such as CLMS \citep{mandic2009complex}, CRLS \citep{livroRLSComplexo}, and LAD \citep{algoritmolad} in a channel equalization problem.
Figure \ref{figuraSistema} shows the block diagram of a communication system with adaptive channel equalization.

\begin{figure}[H]
\centering
\subfloat[Clean input signal $(A_k)$]{\includegraphics[width=3.3in]{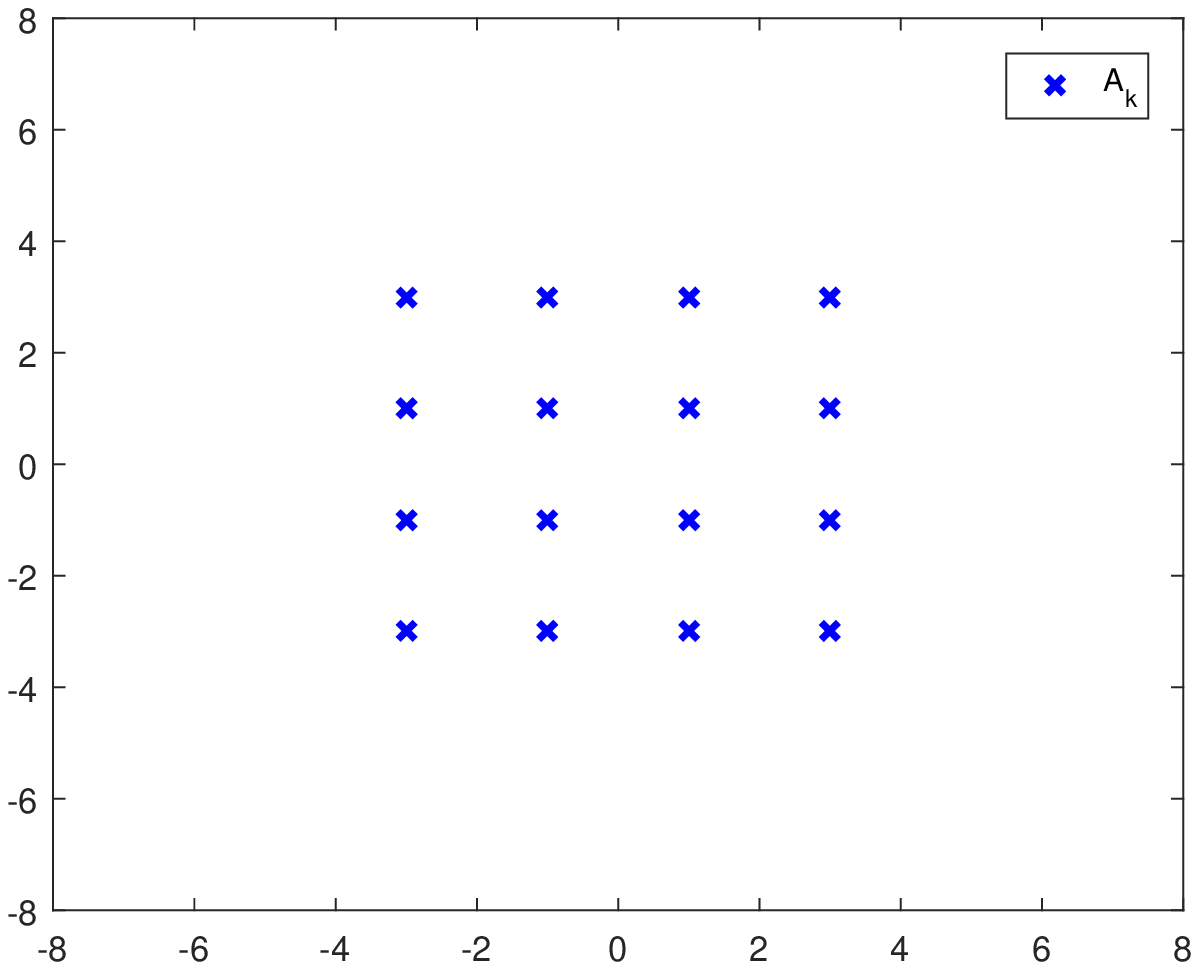}} 
\subfloat[Signal distorted by the Channel $(B_k)$]{\includegraphics[width=3.3in]{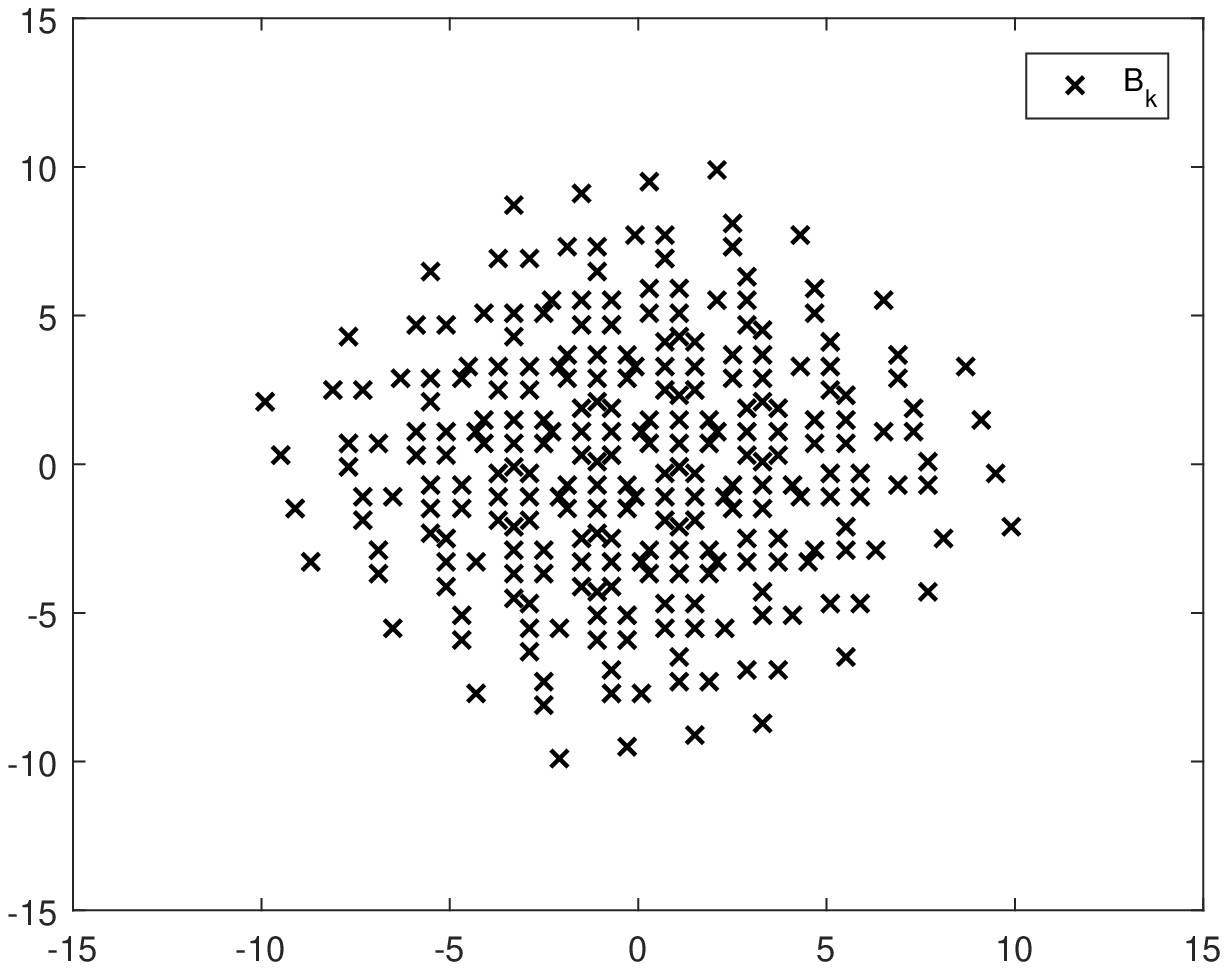}}\\
\subfloat[$B_k$ when corrupted with alpha-stable noise $(C_k)$. Alpha = 1.8 and GSNR = 20dB.]{\includegraphics[width=3.3in]{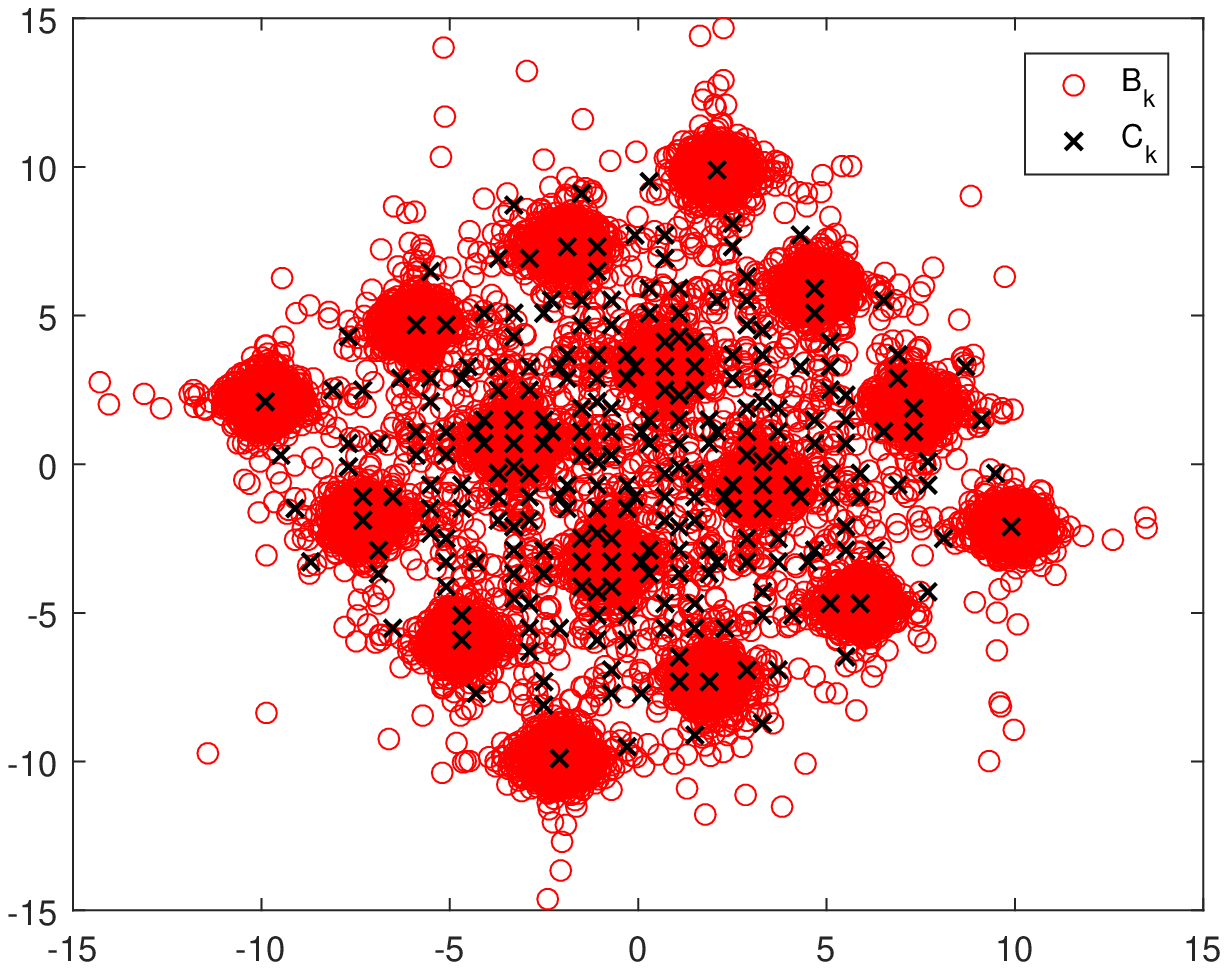}} 
\subfloat[Recovered signal after adaptive equalization using MCCC using kernel size $\sigma = 1$ $(D_k)$]{\includegraphics[width=3.3in]{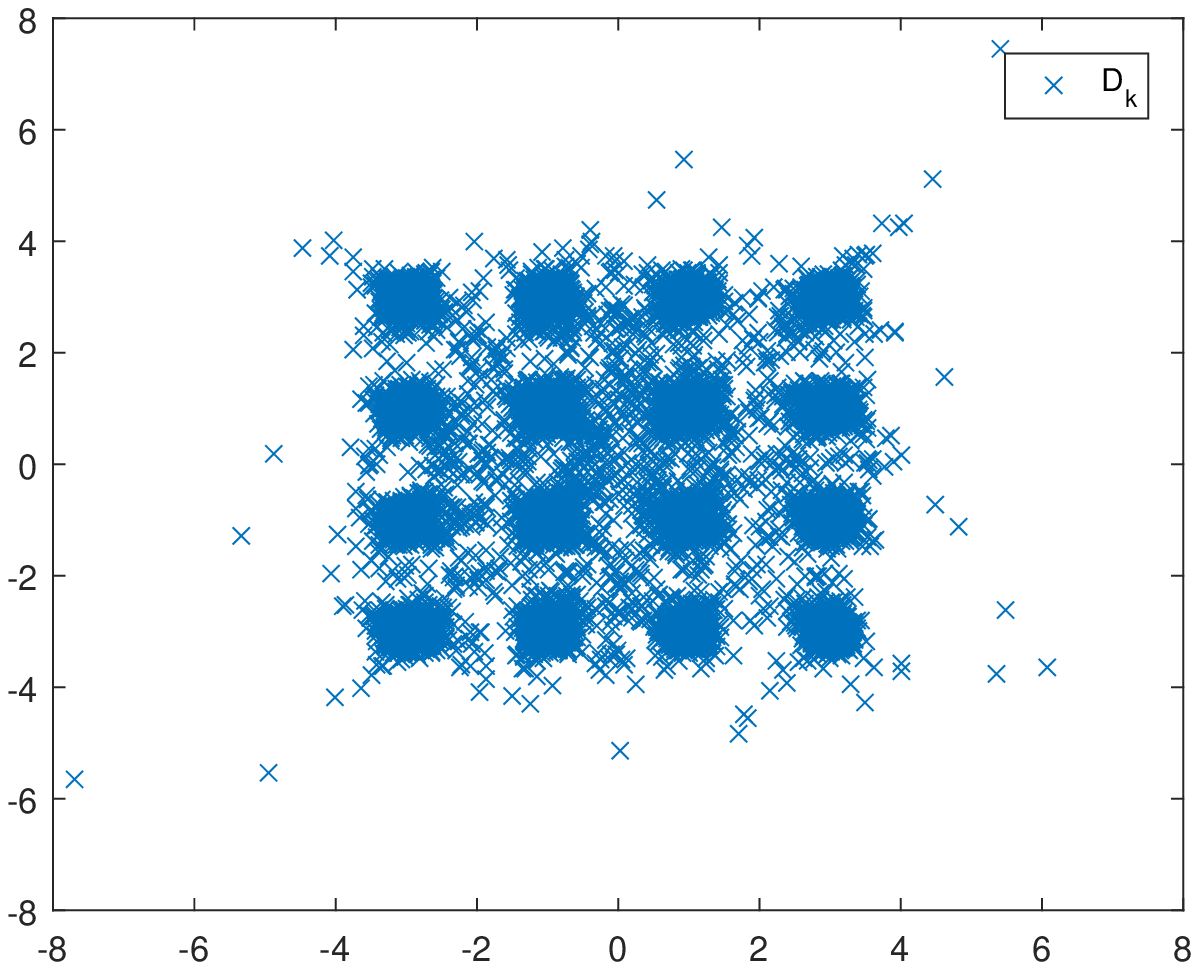}}\\
\caption{Data flow analysis, where each letter represents the corresponding stage in Figure \ref{figuraSistema}.} 
\label{dataflow} 
\end{figure}

The in-phase component $A_{k;I}$ and quadrature component $A_{k;Q}$ at $A$th instant are transmitted through a delay line to obtain the current and past signals. The training signal is formed by a sequence of T$-$spaced complex symbols associated to a $16-QAM$ (Quadrature Amplitude Modulation) constellation whose values $(\pm 1 \pm j1, \pm 2 \pm j2, ... , \pm 8 \pm j8)$ are seen in Figure \ref{dataflow}-a. This signal is then applied to an unknown channel $W$ modeled as $W = [w_1,w_2]^T$, where $w_1$ and $w_2 \in \mathbb{C}$. This generates $B_k$, which is represent as

\begin{equation}
    B_{K} = \begin{bmatrix}
        A_K & A_{K-1}
     \end{bmatrix}
    \begin{bmatrix}
        w_1 \\
        w_2
    \end{bmatrix}
\end{equation}
where $w_1$ and $w_2$ were arbitrary chosen as $(1.1 - j1.1)$ and $(0.9 -j0.2)$, respectively. The distorted signal $B_K$ can be observed in Figure \ref{dataflow}-b.

Thus, in order to assess the robustness of the methods to outliers, $B_K$ is corrupted with an additive noise $\eta_k$ producing signal $C_K = B_K + \eta_K$ as shown in Figure \ref{dataflow}-c, where $\eta_k$ is characterized by $\alpha$-stable distribution with $\alpha=1.8$. The impulsive noise strength is measured by the generalized signal to noise ratio  (GSNR) \citep{medidagsnr} defined as

\begin{equation}
GSNR = 10log_{10}\left ( \frac{1}{\gamma M}\sum_{t=1}^{M}|s(t)^2| \right )
\end{equation}
where $\gamma$ is the dispersion parameter of $\alpha$-stable noise. Different GSNR values varying from 10 dB to 20 dB have been used in the simulation tests. It is worth to mention that the result in Figure 2-c was obtained when GNSR=20 dB.

\begin{figure}[H]
\centering
\subfloat[Bit Error Rate for the tested algorithms.]{\includegraphics[width=3.3in]{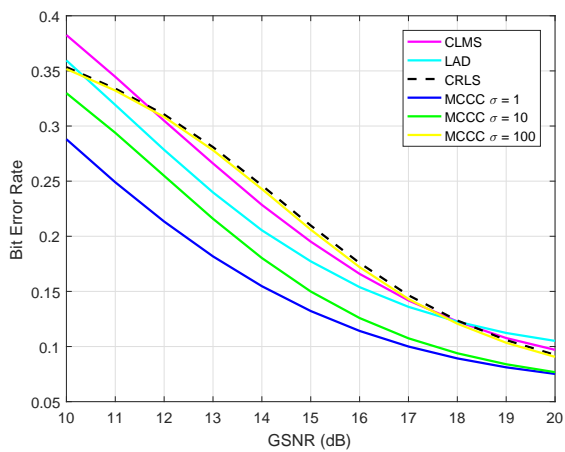}} 
\subfloat[Standard deviation of the Bit Error Rate. ]{\includegraphics[width=3.3in]{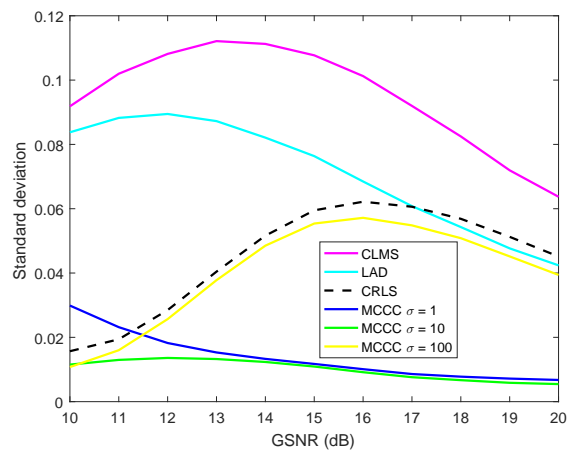}}\\
\caption{BER (left) and respective standard deviation (right) resulting from $10^5$ iterations of the tested algorithms for different GSNR scenarios.}
\label{ber} 
\end{figure}

Signal $C_k$ and its respective delay are used as inputs of the adaptive equalizer, as well as the error between the training signal and the equalizer output $D_K$. In the simulations, $W$ is always initialized with zeros. While MCCC and also CRLS are fixed-point solutions, both CLMS and CMOD  techniques employ an ascendant gradient with learning rate of 0.01. After the equalization process with the training signal, the parameters found by each algorithm at every iteration are used to generate a Bit Error Rate (BER) for all tested GSNR levels. In order to compare the results in Figure \ref{ber}-a with the ones provided by classical methods e.g. CLMS, CRLS, and LAD, three different values of $\sigma$ are chosen i.e. 1, 10, and 100. The performances of algorithms have been evaluated for an average of $10^{5}$ Monte Carlo trials in an enviroment with $alpha=1.5$ and GSNR varing from 10 to 20 db.

\begin{figure}[H]
	\centering
	\includegraphics[width=4.1in]{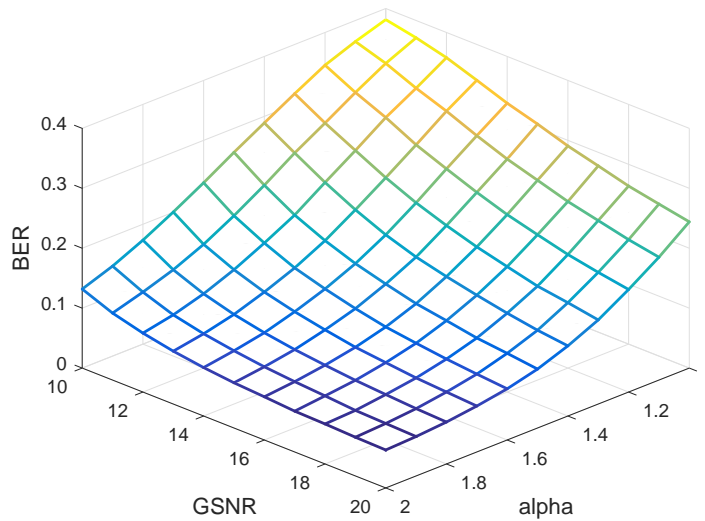}
	\caption{BER for the MCCC algorithm with fixed kernel size $\sigma=1$ for distinct GSNR scenarios and values of alpha.} 
	\label{3d2}
\end{figure}

As it can be seen in Figure \ref{ber}-a, all methods are able to achieve equivalent good results for high values of GSNR. However, as this parameter decreases associated to the nature of the selected type of noise with $\alpha$-stable distribution, the probability of existence of outliers increases. This type of environment degrades the performance level of second-order methods such as CRLS and CLMS. Figure \ref{3d2} shows the BER performance regarding MCCC for a kernel size $\sigma=1$ and different values of alpha, which vary from 1 to 2. According to property 3, when using the Gaussian kernel, the complex correntropy is a weighted sum of all the even moments of the probability density estimation from event $C_1 = C_2$, what makes MCCC robust to outliers similarly to LAD.

\begin{figure}[H]
	\centering
	\includegraphics[width=4.1in]{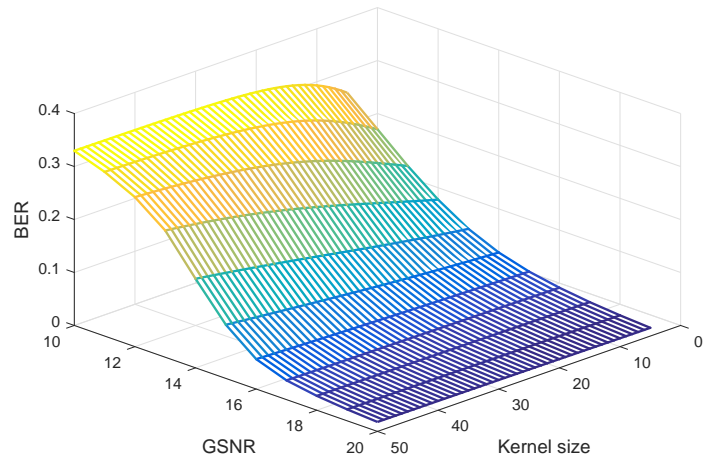}
	\caption{Influence of different kernel sizes on the BER behavior for the MCCC algorithm with different values of GSNR, where $\alpha = 1.5 $ is fixed. } 
	\label{ber3d}
\end{figure}

Additionally, the standard deviation of the Monte Carlo trials in Figure \ref{ber}-a is shown Figure \ref{ber}-b. One can notice that, for all tested kernel sizes and GSNR values, MCCC has the smallest standard deviation among all tested algorithms thus demonstrating its robustness. It is also worth to mention that this method has fast convergence rates since it is a fixed-point method \citep{7763864}. According to Figure \ref{ber}-a, the MCCC performance regarding the channel equalization problem is strictly related to the selection of a proper kernel size, which is a free parameter and must be wisely chosen by the user. The best results could be obtained with a kernel size $\sigma=1$, where Figure \ref{dataflow}-d shows the undistorted signal $D_K$ produced by the MCCC method with $\sigma=1$ when $GSNR=20 dB$. Therefore, one can notice, for example, that the LAD algorithm outperforms MCCC when $\sigma=100$ for some values of GSNR. According to property 4, when the kernel size is increased, complex correntropy tends to behave as the correlation analogously to correntropy in the real-valued case. Then, as expected, one can notice how similar the MCCC performance is to that regarding CRLS when $\sigma=100$. In order to highlight the importance of the kernel size adjustment, Figure \ref{ber3d} is presented to show a surface formed by each BER curve created with different kernel sizes within the range from 5 to 50.

Once the kernel size is adjusted, the MCCC method is able to achieve better BER levels than the classic algorithms for all tested values of GSNR, being an effective tool to deal with outlier environments.

\section{Conclusions}\label{conc}

This paper has expanded the definition of complex correntropy to positive-definite kernels. Furthermore, a complete derivation of this similarity measure is presented as well as important properties such as symmetry, boundedness, high-order statistical measure, probabilistic meaning, mathematical relationship with classical correntropy, and polar coordinate representation. This expanded definition together with the properties makes complex correntropy structurally similar to the correntropy function. For positive valued kernels with complex arguments, complex correntropy could be used as a generalization of conventional real-valued arguments correntropy.

A brief review on Wirtinger calculus has been discussed in order to use the complex correntropy as a cost function in a channel equalization problem, thus resulting in the MCCC algorithm, which was analyzed in detail. Therefore, the advantages associated to the use of complex correntropy in nonGaussian signal processing have been clearly demonstrated both theoretically and experimentally.

The results obtained from the channel equalization problem have shown the improved performance of the complex correntropy function in nonGaussian environments when compared to classic algorithms such as LAD, CLMS, and CRLS. After setting the proper kernel size, the MCCC method has been able to overcome the performance of other similar algorithms, thus providing better values of BER for all tested scenarios. Being a fixed-point solution, it also provides MCCC with a fast convergence rate. There are several potential applications that may explore the benefits from this similarity measure. For instance, spectrum sensing and automatic modulation classification are interesting lines for further research. Another approach lies in understanding complex correntropy when using the theory of reproducing kernel Hilbert space.

\section*{References}

\bibliographystyle{model2-names}\biboptions{authoryear}
\bibliography{referencias}

\newpage
\section*{Appendix I}

\textit{Property 4.1:} Assuming i.i.d data $\{ ( x_i,y_i,z_i,s_i )_{i=1}^N \}$ are draw from the joint PDF $f_{xyzs}$, and $\hat{f_{\sigma}}_{xyzs}$ is its Parzen estimate with kernel size $\sigma$, the complex correntropy estimated with kernel size $\sigma' = \sigma \sqrt{2}$ is the integral of $\hat{f_{\sigma}}_{xyzs}$ along the plane formed by $x=y$ and $z=s$.

\begin{equation}\label{querseprovar}
\hat{V}^{C}_{\sigma'}(C_1,C_2) = \int\limits_{-\infty}^{\infty} \int\limits_{-\infty}^{\infty}  \! \hat{f_{\sigma}}_{XYZS}(x,y,z,s) \, \mathrm{d}u_{1}\mathrm{d}u_{2} \Big|_{x=y=u_{1}, z=s=u_{2}}
\end{equation}

\textit{Proof.}

Lets assume two complex random variables $C_{1}=X+j\,Z$ and  $C_{2} = Y+j\,S$, where $C_{1},C_{2} \in \mathbb{C}$, and $X,Y,Z,S$  are real-valued random variables. Complex correntropy was develop to keep the probabilistic meaning from correntropy. So, estimate the complex correntropy is also estimate the probability density of the event $C_1=C_2$, which implies in estimate the probability density of the event $X=Y$ and $Z=S$. The reasoning is that two complex numbers are equal when their real parts are equal and their imaginary parts are also equal to each other, then:

\begin{equation}\label{integral4grande}
\hat{P}(C_{1} = C_{2}) = \int\limits_{-\infty}^{\infty} \int\limits_{-\infty}^{\infty} \int\limits_{-\infty}^{\infty} \int\limits_{-\infty}^{\infty} \hat{f}_{XYZS}(x,y,z,s) \delta(x-y) \delta(z-s) \mathrm{d}x\mathrm{d}y \mathrm{d}z\mathrm{d}s
\end{equation}

When $x=y$ and $z=s$, equation (\ref{integral4grande}) can be rewritten as:
\small
\begin{equation}\label{p2}
\hat{P}(C_{1} = C_{2}) =\int\limits_{-\infty}^{\infty} \int\limits_{-\infty}^{\infty}  \! \hat{f}_{XYZS}(x,y,z,s) \, \mathrm{d}u_{1}\mathrm{d}u_{2} \Big|_{x=y=u_{1}, z=s=u_{2}} = \int\limits_{-\infty}^{\infty} \int\limits_{-\infty}^{\infty}  \! \hat{f}_{XYZS}(u_{1},u_{1},u_{2},u_{2}) \, \mathrm{d}u_{1}\mathrm{d}u_{2} 
\end{equation}
\normalsize
which is the right side of the equation \ref{querseprovar}.

Lets recall Equation \ref{ccabertaAnexo}, which estimates correntropy from data using a Gaussian Kernel.

\begin{equation}\label{ccabertaAnexo}
\hat{V}^{C}_{\sigma}(C_{1},C_{2}) = \frac{1}{2\pi\sigma^{2}} \frac{1}{N} \sum\limits_{n=1}^N exp \left ( -\frac{(x_{n} - y_{n})^{2} + (z_{n} - s_{n})^{2} }{2\sigma^2} \right ) 
\end{equation}

Then, to complete the proof, one should obtain Equation \ref{ccabertaAnexo} by solving the double integral from Equation \ref{querseprovar}.

First, lets replace $\hat{f}_{XYZS}$ for the Parzen estimator defined as:

\begin{equation}\label{parzenL}
\hat{f}_{X^{1},X^{2},...X^{L}}(x^{1},x^{2},...,x^{L}) = \frac{1}{N}\sum\limits_{n=1}^N \prod \limits_{l=1}^L  G_{\sigma}(x^{l}-x^{l}_{n})
\end{equation}
where
\begin{equation}\nonumber
G_{\sigma}(x) = \frac{1}{\sqrt{2\pi}\sigma}exp \left ( -\frac{x^2}{2\sigma^2} \right )
\end{equation}

Then, using $L=4$ in equation (\ref{parzenL}) and then replacing $\hat{f}_{XYZS}$ in equation (\ref{p2}):

\begin{equation}\label{adupla}
\hat{V}^{C}_{\sigma}(C_{1},C_{2}) = \int\limits_{-\infty}^{\infty} \int\limits_{-\infty}^{\infty}\frac{1}{N}\sum\limits_{n=1}^N  G_{\sigma}(u_{1}-x_{n}) \, G_{\sigma}(u_{1}-y_{n}) \,G_{\sigma}(u_{2}-z_{n}) \, G_{\sigma}(u_{2}-s_{n})  \mathrm{d}u_{1} \mathrm{d}u_{2} 
\end{equation}

The best way to solve this double integral is rewrite equation (\ref{adupla}) is make:

\begin{equation}\nonumber
\hat{V}^{C}_{\sigma}(C_{1},C_{2}) = \int\limits_{-\infty}^{\infty}\int\limits_{-\infty}^{\infty} \frac{1}{N} \sum\limits_{i=1}^N \frac{1}{(4\pi^{2}\sigma^{4})} exp \left ( - \frac{1}{2\sigma^{2}} ( a(u_1-b)^{2}+ a'(u_2-b')^{2} + c  \right ) \mathrm{d}u_1\mathrm{d}u_2
\end{equation}

and then

\small
\begin{equation}\label{aintegralmae}
\hat{V}^{C}_{\sigma}(C_{1},C_{2}) = \int\limits_{-\infty}^{\infty}\int\limits_{-\infty}^{\infty} \frac{1}{N} \sum\limits_{i=1}^N \frac{1}{(4\pi^{2}\sigma^{4})} exp \left ( - \frac{1}{2\sigma^{2}}  a(u_1-b)^{2} \right) exp \left( - \frac{1}{2\sigma^{2}} a'(u_2-b')^{2} \right)  exp \left( - \frac{c}{2\sigma^{2}} \right ) \mathrm{d}u_1\mathrm{d}u_2
\end{equation}
\normalsize

The coefficients $a,b,a',b'$ and $c$ are obtained by the completing square technique, detailed	 as follows:

\begin{equation}\label{igualdade}
(u_1-x_{i})^{2} + (u_1-y_{i})^{2} + (u_2-z_{i})^{2} + (u_2-s_{i})^{2} = a(u_1-b)^2 + a'(u_2-b')^2 + c 
\end{equation}

Developing the right side of the equation (\ref{igualdade}): 
\begin{equation}\nonumber
au_1^2 + a'u_2^2 - 2u_1ab - 2u_2a'b' + ab^2 + a'b'^2 + c 
\end{equation}

And then, the left side:
\begin{equation}\nonumber
2u_1^2 + 2u_2^2 + x_{i}^2 + y_{i}^2 + z_{i}^2 + s_{i}^2 - 2u_1(x_{i}+y_{i}) - 2u_2(z_{i} + s_{i})
\end{equation}

One can say that:

\begin{equation}\nonumber
a = a' = 2 \text{; }
b = \frac{(x_{i}+y_{i})}{2} \text{; }
b' = \frac{(z_{i}+s_{i})}{2} \text{; and }
c = \frac{(x_{i}-y_{i})^2 + (z_{i}-s_{i})^2}{2}
\end{equation}

Using these values in the equation (\ref{aintegralmae}) we have:

\begin{equation}\label{aint}
= \int\limits_{-\infty}^{\infty}\int\limits_{-\infty}^{\infty} \frac{1}{N}\sum\limits_{i=1}^N \frac{1}{(4\pi^{2}\sigma^{4})} exp \left ( - \frac{(u_1-b)^{2}}{\sigma^{2}}   \right) exp \left( - \frac{(u_2-b')^{2}}{\sigma^{2}}  \right)  exp \left( - \frac{c}{2\sigma^{2}} \right ) \mathrm{d}u_1\mathrm{d}u_2
\end{equation}

Making $\sigma^2 = 2\theta^2$, one can rewrite equation (\ref{aint}) as: 

\begin{small}
\begin{equation}\label{gigante}
=\int\int \frac{1}{N}\sum\limits_{i=1}^N \left ( \frac{1}{2\pi\theta^2} \right ) \left ( \frac{1}{2\pi4\theta^2} \right ) exp \left ( - \frac{(u_1-b)^{2}}{2\theta^2}   \right) exp \left( - \frac{(u_2-b')^{2}}{2\theta^{2}}  \right)  exp \left( - \frac{c}{4\theta^2} \right ) \mathrm{d}u_1\mathrm{d}u_2
\end{equation}
\end{small}

But
\begin{equation}\label{integralFechada2}
 \left (\frac{1}{\sqrt{2\pi}\theta} \right )  \int\limits_{-\infty}^{\infty} exp \left ( - \frac{ (u_1-b)^2}{2\theta^2} \right)\mathrm{d}u_1 \left (\frac{1}{\sqrt{2\pi}\theta} \right ) \int\limits_{-\infty}^{\infty}  exp \left ( - \frac{(u_2-b')^2}{2\theta^2}  \right) \mathrm{d}u_2 = 1
\end{equation}

Then, equation (\ref{gigante}) can be rewrite as: 

\begin{equation}\nonumber
\hat{V}^{C}_{\sigma}(C_{1},C_{2}) = \left ( \frac{1}{2\pi4\theta^2} \right )  \frac{1}{N}\sum\limits_{i=1}^N  exp  \left( - \frac{(x_{i}-y_{i})^2 + (z_{i}-s_{i})^2}{8\theta^2} \right )  
\end{equation}

Or, making $8\theta^2 = (2\sigma') ^2$:

\begin{equation}\label{final}
\hat{V}^{C}_{\sigma}(C_{1},C_{2}) = \left ( \frac{1}{2\pi(\sigma')^2} \right )  \frac{1}{N}\sum\limits_{i=1}^N  exp \left( - \frac{(x_{i}-y_{i})^2 + (z_{i}-s_{i})^2}{2(\sigma')^2} \right ) 
\end{equation}

Where $\sigma' = \sigma\sqrt2$, which implies

\begin{equation}\nonumber
V^{C}_{\sigma}(C_{1},C_{2}) = \frac{1}{N}\sum\limits_{n=1}^N  G_{\sigma \sqrt{2}}(x_{n} - y_{n}) \,G_{\sigma \sqrt{2} }(z_{n} - s_{n}) 
\end{equation}

The non parametric estimator of complex correntropy with Parzen windows can be written as
\begin{equation}\label{finalabertaanexo}
\hat{V}^{C}_{\sigma}(C_{1},C_{2}) = E[G_{\sigma}^{C}(C_1 - C_2)]
\end{equation}

where
\begin{equation}\label{correntropiacomplexa}
G^{C}_{\sigma} (C_1 - C_2 )= \frac{1}{2\pi\sigma^2}exp \left ( -\frac{(C_{1} - C_{2}) (C_{1} - C_{2})^{*}}{2\sigma^2} \right ) = \frac{1}{2\pi\sigma^2} exp \left ( -\frac{(x_{n} - y_{n})^{2} + (z_{n} - s_{n})^{2} }{2\sigma^2} \right ) 
\end{equation}

alternatively,
\begin{equation}\label{finalabertaanexo2}
\hat{V}^{C}_{\sigma}(C_{1},C_{2}) = \frac{1}{2\pi\sigma^{2}} \frac{1}{N} \sum\limits_{n=1}^N exp \left ( -\frac{(x_{n} - y_{n})^{2} + (z_{n} - s_{n})^{2} }{2\sigma^2} \right ) 
\end{equation}
which completes the proof.
\begin{flushright}
$\blacksquare$
\end{flushright}

\section*{Appendix II}
\textit{MCCC Fixed Point Solution:}

Considering a linear model and define the error, $e = d - y$, as the difference between the desired signal $d$ and the filter output $y = \textbf{w}^H \textbf{X}$, with $y$, $d$, $e$, $\textbf{X}$, $\textbf{w} \in \mathbb{C}$, let the new criteria MCCC be defined as the maximum complex correntropy between two random complex variables $D$ and $Y$:

\begin{equation}\nonumber
J_{MCCC} = V^{C}_{\sigma}(D, Y) = E_{DY}[G^{C}_{\sigma\,\sqrt{2}}(D-\textbf{w}^{H}\textbf{X})] =  E_{DY}[G^{C}_{\sigma\,\sqrt{2}}(e)]
\end{equation}

For archiving the FP solution for the optimal weights, one can set the cost function derivative to zero in respect to $\textbf{w}^*$ 

\begin{equation}\nonumber
\frac{\partial J_{MCCC} }{\partial w^{*}} = \frac{\partial E_{DY}[G^{C}_{\sigma\,\sqrt{2}}(e)] }{\partial w^{*}}  = E_{DY} \left [ G^{C}_{\sigma\,\sqrt{2}}(e)\frac{\partial (ee^*) }{\partial w^{*}} \right ]   =  \textbf{0}
\end{equation}

\begin{equation}\nonumber
\frac{\partial (ee^*)}{\partial w^*} = \frac{\partial (D-\textbf{w}^{H}\textbf{X})(D-\textbf{w}^{H}\textbf{X})^* }{\partial w^*}
\end{equation}

\begin{equation}\nonumber
\frac{\partial (ee^*)}{\partial w^*} = \frac{\partial (D-\textbf{w}^{H}\textbf{X})(D^*-\textbf{w}^{T}\textbf{X}^*) }{\partial w^*}
\end{equation}

Then, lets find the derivative of $ee^*$ in respect to $w^{*}$ using Wirtinger Calculus:

\begin{equation}\nonumber
\frac{\partial (ee^*)}{\partial w^*} = \frac{\partial (DD^* - D\textbf{X}^{H}w - \textbf{w}^{H}\textbf{XX}^{*} + \textbf{w}^{H}\textbf{XX}^{H}\textbf{w} ) }{\partial w^{*}}
\end{equation}

\begin{equation}\nonumber
\frac{\partial DD^* }{\partial \textbf{w}^*} = 0
\end{equation}
\begin{equation}\nonumber
\frac{\partial D\textbf{X}^{H}w }{\partial \textbf{w}^*} = 0
\end{equation}
\begin{equation}\nonumber
\frac{\partial \textbf{w}^{H}\textbf{X}D^{*} }{\partial \textbf{w}^*} = \frac{\partial \textbf{X}^T \textbf{w}^*D^{*} }{\partial \textbf{w}^*} = \textbf{X}D^*
\end{equation}
\begin{equation}\nonumber
\frac{\partial \textbf{w}^{H}\textbf{XX}^{H}\textbf{w} }{\partial \textbf{w}^*} = \frac{\partial \textbf{w}^T \textbf{X}^* \textbf{X}^T \textbf{w}^*  }{\partial \textbf{w}^*} = \textbf{X}\textbf{X}^H \textbf{w}
\end{equation}
always using the denominator layout ( result is column vector) 

\begin{equation}\nonumber
\frac{\partial (ee^*)}{\partial w^*} = (\textbf{X}D^* - \textbf{XX}^{H}\textbf{w})
\end{equation}

\begin{equation}\nonumber
E_{DY}[G^{C}_{\sigma}(e) (\textbf{X}D^* - \textbf{XX}^{H}\textbf{w}) ] = \textbf{0}
\end{equation}

\begin{equation}\nonumber
E_{DY}[G^{C}_{\sigma}(e)\textbf{X}\,D^{*}] = E_{DY}[G^{C}_{\sigma}(e)\textbf{XX}^{H}]\,\textbf{w}
\end{equation}

\begin{equation}
\textbf{w} = \left [ \sum_{n=1}^{N} G^{C}_{\sigma}(e_{n})\textbf{X}_{n}\textbf{X}_{n}^{H} \right ]^{-1} \left [ \sum_{n=1}^{N} G^{C}_{\sigma}(e_{n})\textit{D}_{n}^{*}\,\textbf{X}_{n} \right ]
\end{equation}

A full discussion on how to implement the above equation is presented in \citep{7763864}.







\end{document}

%% file: teste.tex
\begin{picture}(0,0)%
\includegraphics{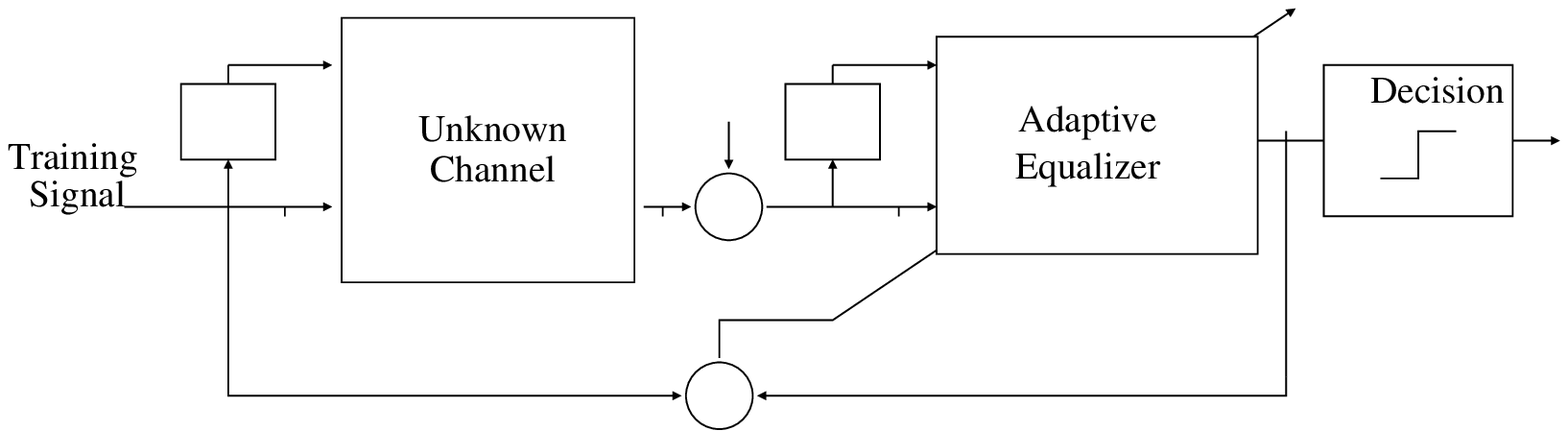}%
\end{picture}%
\setlength{\unitlength}{4144sp}%
\begingroup\makeatletter\ifx\SetFigFont\undefined%
\gdef\SetFigFont#1#2#3#4#5{%
  \reset@font\fontsize{#1}{#2pt}%
  \fontfamily{#3}\fontseries{#4}\fontshape{#5}%
  \selectfont}%
\fi\endgroup%
\begin{picture}(7508,2024)(1370,-2748)
\put(4861,-2626){\makebox(0,0)[b]{\smash{{\SetFigFont{10}{12.0}{\familydefault}{\mddefault}{\updefault}{\color[rgb]{0,0,0}$\sum$}%
}}}}
\put(4906,-1726){\makebox(0,0)[b]{\smash{{\SetFigFont{10}{12.0}{\familydefault}{\mddefault}{\updefault}{\color[rgb]{0,0,0}$\sum$}%
}}}}
\put(4816,-1231){\makebox(0,0)[b]{\smash{{\SetFigFont{10}{12.0}{\familydefault}{\mddefault}{\updefault}{\color[rgb]{0,0,0}noise $\eta_n$}%
}}}}
\put(5401,-1321){\makebox(0,0)[b]{\smash{{\SetFigFont{10}{12.0}{\familydefault}{\mddefault}{\updefault}{\color[rgb]{0,0,0}$z^{-1}$}%
}}}}
\put(2521,-1321){\makebox(0,0)[b]{\smash{{\SetFigFont{10}{12.0}{\familydefault}{\mddefault}{\updefault}{\color[rgb]{0,0,0}$z^{-1}$}%
}}}}
\put(5176,-2491){\makebox(0,0)[b]{\smash{{\SetFigFont{8}{9.6}{\familydefault}{\mddefault}{\updefault}{\color[rgb]{0,0,0}$+$}%
}}}}
\put(4591,-2491){\makebox(0,0)[b]{\smash{{\SetFigFont{8}{9.6}{\familydefault}{\mddefault}{\updefault}{\color[rgb]{0,0,0}$-$}%
}}}}
\put(4591,-1861){\makebox(0,0)[b]{\smash{{\SetFigFont{10}{12.0}{\familydefault}{\mddefault}{\updefault}{\color[rgb]{0,0,0}B}%
}}}}
\put(5716,-1861){\makebox(0,0)[b]{\smash{{\SetFigFont{10}{12.0}{\familydefault}{\mddefault}{\updefault}{\color[rgb]{0,0,0}C}%
}}}}
\put(2791,-1861){\makebox(0,0)[b]{\smash{{\SetFigFont{10}{12.0}{\familydefault}{\mddefault}{\updefault}{\color[rgb]{0,0,0}A}%
}}}}
\put(7561,-1276){\makebox(0,0)[b]{\smash{{\SetFigFont{10}{12.0}{\familydefault}{\mddefault}{\updefault}{\color[rgb]{0,0,0}D}%
}}}}
\put(4996,-1411){\makebox(0,0)[b]{\smash{{\SetFigFont{8}{9.6}{\familydefault}{\mddefault}{\updefault}{\color[rgb]{0,0,0}$+$}%
}}}}
\put(4636,-1591){\makebox(0,0)[b]{\smash{{\SetFigFont{8}{9.6}{\familydefault}{\mddefault}{\updefault}{\color[rgb]{0,0,0}$+$}%
}}}}
\end{picture}%